\documentclass[10pt,conference]{IEEEtran}
\usepackage{amssymb,amsmath,palatino}

\begin{document}
\title{Efficient evaluations of polynomials\\  over finite fields}
\author{
\authorblockN{Davide Schipani}
\authorblockA{Mathematics Institute\\
University of Z\"urich\\
CH-8057 Z\"urich\\
davide.schipani@math.uzh.ch}
\and
\authorblockN{Michele Elia}
\authorblockA{Dipartimento di Elettronica\\
 Politecnico di Torino\\
 IT-10129 Torino \\
elia@polito.it}
\and
\authorblockN{Joachim Rosenthal}
\authorblockA{Mathematics Institute\\
University of Z\"urich\\
CH-8057 Z\"urich\\
http://www.math.uzh.ch/aa}

}
\maketitle

\begin{abstract}
\noindent
A method is described which allows to evaluate efficiently a
polynomial in a (possibly trivial) extension of the finite field of
its coefficients. Its complexity is shown to be lower than that of
standard techniques when the degree of the polynomial is large with
respect to the base field. Applications to the syndrome computation in
the decoding of cyclic codes, Reed-Solomon codes in particular, are
highlighted.
\end{abstract}

\vspace{2mm}

\vspace{2mm}


\vspace{8mm}

\section{Introduction}
Standard algorithms for decoding Reed-Solomon and BCH codes such as
the Peterson-Gorenstein-Zierler algorithm involve the evaluation of
polynomials at several steps. In particular instances the complexity
of the algorithms are even dominated by that task~\cite{vetterli}. In
this paper we propose a new method to perform the evaluation efficiently.

The standard technique to evaluate polynomials over a field is
Horner's rule (e.g. \cite[p.467]{knuth2}), which computes the value $P(\alpha)$ for a polynomial
$P(x)=a_n x^n+a_{n-1} x^{n-1}\cdots+a_0$ in an iterative way as
suggested by the following description
$$
(\cdots((a_n\alpha+a_{n-1})\alpha+a_{n-2})\alpha+\cdots)\alpha+a_1)\alpha
+a_0 ~~.
$$
This method requires $n$ multiplications and $n$ additions. In the
following we describe another method to evaluate polynomials with
coefficients over a finite field $GF(p^m)$ and we estimate its
complexity. For that we consider, as is customary, just the number
of multiplications, as in $GF(2^m)$ to multiply is more expensive than
to add: the cost of an addition is $O(m)$ in space and $1$ clock in
time, while the cost of a multiplication is $O(m^2)$ in space and
$O(\log_2 m)$ in time \cite{elia}. We keep track of the number
of additions, too, to be sure that a reduction in the number of
multiplications does not come together with an exorbitant increase in
the number of additions.

Our approach exploits the Frobenius automorphism and its group
properties, therefore we call it polynomial automorphic evaluation.

 \section{Polynomial automorphic evaluation}

 Consider a finite field $GF(q)$ of cardinality $q=p^m$, $p$ a prime, a
 polynomial $P(x)$ of degree $n$, and let $\alpha$ denote an element
 of $GF(q)$. We write $P(x)$ as
$$
P_0(x^p)+ x P_1(x^p)\cdots +x^{p-1} P_{p-1}(x^p) ~~,
$$
where $P_0(x^p)$ collects the powers of $x$ with exponent a multiple
of $p$ and in general $x^{i} P_{i}(x^p)$ collects the powers of the form
$x^{ap+i}$, with $a\in\mathbb{N}$ and $0\leq i\leq p-1$ (see some examples in the following remarks).

If $\sigma$ is the Frobenius automorphism of $GF(p^m)$ mapping $a$ to
$a^p$, we can write the expression above as
$$
P_0^{-1}(x)^p+ x P_1^{-1}(x)^p\cdots +x^{p-1} P_{p-1}^{-1}(x)^p ~~,
$$
where $P_i^{-k}(x)$ stands for the polynomial obtained from $P_i(x)$
by substituting its coefficients with their transforms through
$\sigma^{-k}$, for any $k$ in the set $\{1,\ldots,m\}$. Notice that the polynomials $P_i^{-1}(x)$ have degree
at most $\frac{n-i}{p}$. We can take the exponent out of the brackets
as the field has characteristic $p$.

$P(\alpha)$ for a particular value $\alpha$ can be then obtained from
$\{P_i^{-1}(\alpha) \}$ by making $p$ $p$-th powers, $p-1$
multiplications and $p-1$ sums.

The procedure can be iterated until the polynomials we obtain have
small degree: at each step the number of polynomials is multiplied by
$p$ and their degree is divided roughly by $p$.  For each step we have
to compute $N$ $p$-th powers, where $N$ is the number of polynomials
at that step, while additions and multiplications are slightly less,
as computed below.

If we perform $L$ steps, we have $p^L$ polynomials of degree nearly
$\frac{n}{p^L}$ and the total cost of evaluating $P(\alpha)$
comprehends the following:
\begin{itemize}
\item Evaluation of $p^L$ polynomials of degree $\frac{n}{p^L}$ in
  $\alpha$
\item Computation of $p+p^2+ \cdots + p^L=\frac{p^{L+1}-p}{p-1}$
  $p$-th powers.
\item Computation of $p-1+(p^2-p)+\cdots + p^L-p^{L-1}=p^{L}-1$
  multiplications by powers of $\alpha$.
\item Computation of $p-1+(p^2-p)+\cdots + p^L-p^{L-1}=p^{L}-1$
  additions.
\item Computation of the coefficients of the $p^L$ polynomials through
  $\sigma^{-L}$; the number of coefficients is the same as the number
  of coefficients of $P(x)$, that is at most $n+1$, which would
  possibly imply too many multiplications.  However, we can spare a
  lot, if we do the following: we evaluate the $p^L$ polynomials in
  $\sigma^L(\alpha)$ and then we apply $\sigma^{-L}$ to the outputs.
  So we need to apply powers of $\sigma$ a number of times not greater
  than $p^L+1$. Notice also that what matters in $\sigma^L$ is $L$
  modulo $m$ because $\sigma^m$ is the identity automorphism.
\end{itemize}

So all together we would like to minimize the following number of
multiplications:
\begin{multline*}
  G(L)= 2\lfloor \log_2 p \rfloor \frac{p^{L+1}-p}{p-1}+p^{L}-1+\\
  2\lfloor \log_2 p \rfloor(m-1)(p^L+1)+\frac{n}{p^L}(p^m-1) ~~,
\end{multline*}
where $2\lfloor \log_2 p \rfloor$ refers to a $p$-th power made by
successive squaring (this upper bound is substituted by $1$ when $p$
is $2$), the automorphism $\sigma^L$ counts like a power with exponent
$p^L$, with $L\leq m-1$, and $\frac{n}{p^L}$ are the powers of
$\alpha$ we need to compute, while $p^m-1$ are all their possible
nonzero coefficients. Once we have the powers of $\alpha$ multiplied
by the possible coefficients, we actually need also to compute at most
$n$ additions to get the value of the polynomials.

Since $G(L)$ is a sum of two positive functions, the first
monotonically decreasing and the second increasing with $L$, the
minimum of $G(L)$, considered as a continuous function of $L$, is
unique. A very good estimation of the minimum is then obtained by
computing the derivative of $G(L)$ with respect to $L$, so that the
optimum $L$ is roughly
\begin{equation}
  \label{optpoint}
  \log_p\left(\frac{\sqrt{n(p^m-1)}}{\sqrt{1+2\lfloor \log_2 p 
\rfloor(m-1+\frac{p}{p-1})}}\right) ~~.
\end{equation}
The corresponding minimum can be written as
\begin{multline}
  \label{optvalue}
  2\sqrt{n(p^m-1)}\sqrt{1+2\lfloor \log_2 p \rfloor(m-1+\frac{p}{p-1})}+\\
  2\lfloor \log_2 p \rfloor (m-1)-1- \frac{2\lfloor \log_2 p \rfloor
    p}{p-1} ~~.
\end{multline}
This brings a total cost less than $n$ (Horner's cost) whenever $p^m$
is not too big with respect to $n$.

\paragraph{Remark 1.}
If the coefficients are known to belong to $GF(p)$, then the total
cost is at most
$$
2\lfloor \log_2 p \rfloor
\frac{p^{L+1}-p}{p-1}+p^{L}-1+\frac{n}{p^L}(p-1) ~~,
$$
since $\sigma$ does not change the coefficients in this case. Then the
best value for $L$ is approximately
$$
\log_p\left(\frac{\sqrt{n(p-1)}}{\sqrt{1+2\lfloor \log_2 p
      \rfloor\frac{p}{p-1}}}\right) ~~,
$$
and the total cost becomes even more appealing, in particular when
$p=2$ it is less than $2\sqrt{3n}$.

In this case every step is very straightforward: the decomposition of a polynomial $P(x)$ as a sum of two polynomials by collecting odd and even powers
 of $x$ is
$$  P(x) = P_0(x^2)+ x P_1(x^2)= P_0(x)^2 + xP_1(x)^2  ~~. $$

Actually this case happens often in coding theory \cite{sloane}, in
particular in the computation of syndromes for a binary code. 
In this situation we can have as additional advantage the possibility of
precomputing the powers of $\alpha$, since what is usually needed is
to evaluate a polynomial in several powers of a particular value
$\alpha$.

\paragraph{Remark 2.}

Similarly, if the coefficients belong to $GF(p^d)$ for a divisor $d$
of $m$, then the total cost is at most
\begin{multline*}
  2\lfloor \log_2 p \rfloor \frac{p^{L+1}-p}{p-1}+p^{L}-1\\
  +2\lfloor \log_2 p \rfloor(d-1)(p^L+1)+\frac{n}{p^L}(p^d-1) ~~.
\end{multline*}
And the best value for $L$ is
$$
\log_p\left(\frac{\sqrt{n(p^d-1)}}{\sqrt{1+2\lfloor \log_2 p
      \rfloor(d-1+\frac{p}{p-1})}}\right) ~~.
$$

\paragraph{Remark 3.}

If $p^m \approx n$, i.e. $m \approx \frac{\log_2 n}{\log_2 p}$, which
is the case of the Reed-Solomon codes, the proposed method does not
seem to give any advantage as the complexity is approximately $2 n
\sqrt{2\log_2 n} > n$ by Equation (\ref{optpoint}) .  However, if $m$
is not prime, then a gain is still possible, by using the previous
remarks. Let us show an example below.  Suppose $m$ is even. Then the
elements of the field $GF(p^m)$ can be represented in the form $a+b
\beta$, where $a,b \in GF(p^{m/2})$ and $\beta$ is a root of a
quadratic polynomial irreducible over
$GF(p^{m/2})$.
Therefore, the polynomial $p(x)$ with coefficients in $GF(p^m)$ can be
written as a sum $p_1(x)+ \beta p_2(x)$ where both $p_1(x)$ and
$p_2(x)$ have coefficients in $GF(p^{m/2})$: if we evaluate these two
polynomials using the proposed algorithm, the cost for each evaluation
is
\begin{multline*}
2\sqrt{np^{m/2}}\sqrt{1+2\lfloor \log_2 p \rfloor(m/2-1+\frac{p}{p-1})}\\
+2\lfloor \log_2 p \rfloor (m/2-1)-1-
 \frac{2\lfloor \log_2 p \rfloor p}{p-1} ~~, 
\end{multline*}
and to get the total cost we multiply by $2$. For example, if $p=2$
and $2^m \approx n$, the total cost is approximately $2\sqrt{2} \sqrt[4]{n^3}
\sqrt{\log_2 n} $, a figure significantly less than $n$ when $m > 12$.

 \paragraph{Remark 4.}
 Given the importance of cyclic codes over $GF(2^m)$, for instance the
 Reed-Solomon codes that are used in any CD rom, or the famous
 Reed-Solomon code $[255,223,33]$ over $GF(2^{8})$ used by NASA (\cite{wicker2}), an
 efficient evaluation of polynomials over $GF(2^m)$ in points of the
 same field is of the greatest interest.

 In the previous remarks, we have shown that non-trivial gains are
 possible, however, in particular scenarios an additional gain can be
 obtained by choosing $L$ as a factor of $m$ which is close to the
 value obtained in equation (\ref{optpoint}), together with some
 arrangements as explained below.

 The idea will be illustrated considering the decoding of the above
 mentioned Reed-Solomon code. We will only show how to obtain the $32$
 syndromes; the decoding is done from that point on using the standard
 Berlekamp-Massey algorithm, the Chien search to locate the errors,
 and the Forney algorithm to compute the error
 magnitudes~\cite{blahut}.

 Let $r(x)=\sum_{i=0}^{254} r_i x^i$, $r_i \in GF(2^8)$, be a received
 code word of a Reed Solomon code $[255,223,33]$ generated by the
 polynomial $g(x)=\prod_{i=1}^{32}(x-\alpha^i)$, with $\alpha$ a
 primitive element of $GF(2^8)$, i.e. a root of $x^8+x^5+x^3+x+1$. Our
 aim is to evaluate the syndromes $S_j=r(\alpha^j)$, $j=1, \ldots ,
 32$.

 \noindent We can argue in the following way.  The power
 $\beta=\alpha^{17}$ is a primitive element of the subfield $GF(2^4)$,
 it is a root of the polynomial $x^4+x^3+1$, and has trace $1$ in
 $GF(2^4)$. Therefore, a root $\gamma$ of $z^2+z+\beta$ is not in
 $GF(2^4)$ (see \cite[Corollary 3.79, p.118]{lidl}), but it is an
 element of $GF(2^8)$, and every element of $GF(2^8)$ can be written
 as $a+b \gamma$ with $a,b \in GF(2^4)$.
 Consequently, we can write $r(x)=r_1(x)+\gamma r_2(x)$ as a sum of
 two polynomials over $GF(2^4)$, evaluate each $r_i(x)$ in the roots
 $\alpha^j$ of $g(x)$, and obtain each syndrome
 $S_j=r(\alpha^j)=r_1(\alpha^j)+\gamma r_2(\alpha^j)$ with $1$
 multiplication and $1$ sum.

 Now, following our proposed scheme, if $p(x)$ is either $r_1(x)$ or
 $r_2(x)$, in order to evaluate $p(\alpha^j)$ we consider the
 decomposition
 \begin{multline*}
p(x) = (p_0+p_2x+\cdots +p_{254}x^{127})^2 \\
+ x (p_1+p_3x+\cdots +p_{253}x^{126})^2~~,  
\end{multline*}
where we have not changed the coefficients computing $\sigma^{-1}$ for
each of them, as a convenient Frobenius automorphism will come into
play later.
Now, each of the two parts can be decomposed again into the sum of two
polynomials of degree at most $63$, for instance
\begin{multline*}
  p_0+p_2x+\cdots +p_{254}x^{127}\\ = (p_0+p_4x+\cdots
  +p_{252}x^{63})^2
  \\
  + x (p_2+p_6x+\cdots +p_{254}x^{63})^2
\end{multline*}
and at this stage we have four polynomials to be evaluated.  The next
two steps double the number of polynomials and half their degrees; we
write just one polynomial per each stage as an example
\begin{multline*}
  p_0+p_4x+\cdots +p_{252}x^{63} \\
  = (p_0+p_8x+\cdots +p_{248}x^{31})^2 \\
  + x(p_4+p_{12}x+\cdots +p_{252}x^{31})^2
\end{multline*}
\begin{multline*}
p_0+p_8x+\cdots +p_{248}x^{31}\\
 =(p_0+p_{16}x+\cdots +p_{240}x^{15} )^2 \\
+x(p_8+p_{24}x+\cdots +p_{248}x^{15})^2  
\end{multline*}

Since we choose to stop the decomposition at this stage, we have to
evaluate $16$ polynomials of degree at most $15$ with coefficients in
$GF(16)$, but before doing this computation we should perform the
inverse Frobenius automorphism $\sigma^{-4}$ on the coefficients,
however $\sigma^{-4}(p_i)=p_i$ because the coefficients are in
$GF(16)$ and any element $\beta$ in this field satisfies the condition
$\beta^{2^4}=\beta$.

Now, let $K$ be the number of code words to be decoded. It is
convenient to compute only once the following field elements:
%

\begin{itemize}
\item $\alpha^i$, $i= 2, \ldots , 254$ and this requires $253$
  multiplications;
\item $\alpha^i \cdot \beta^j$ for $i=0,\ldots ,254$ and $j=1, \ldots,
  14$, which requires $255 \cdot 14=3570$ multiplications.
\end{itemize}
Then only sums (that can be performed in parallel) are required to
evaluate $16$ polynomials of degree $15$ for each $\alpha^j$, $j=1
\ldots , 32$. Once we have the values of these polynomials, in order
to reconstruct each of $r_1(\alpha^j)$ and $r_2(\alpha^j)$, we need
\begin{itemize}
\item $16+8+4+2$ squares
\item $8+4+2+1$ multiplications (and the same number of sums).
\end{itemize}
Summing up, every $r(\alpha^j)=r_1(\alpha^j)+\gamma r_2(\alpha^j)$ is
obtained with $2 \cdot 45+1=91$ multiplications.  Then the total cost
of the computation of $32$ syndromes drops down from $31+32 \cdot
254=8159$ with Horner's rule to $32 \cdot 91+3570+253 =6735$. Since we
have $K$ code words the total cost drops from $31+8128 \cdot K$ to $3823+
2912 \cdot K$, with two further advantages:

- many operations can be parallelized, so that the speed is further
increased;

- the multiplications can be performed in $GF(2^4)$ instead of
$GF(2^8)$, if we write $\alpha^j=a_j+\gamma b_j$; the number of
multiplications could increase but their speed would be much faster.

Clearly, these decoding schemes can be generalized for cyclic codes
over any $GF(p^m)$ with $m$ not prime.

\section*{Acknowledgment}
The Research was supported in part by the Swiss National Science
Foundation under grant No. 126948



\begin{thebibliography}{99}
\bibitem{blahut} R.E. Blahut, {\em Theory and Practice of Error
    Control Codes}, Addison-Wesley, Reading Massachussetts, 1983.
\bibitem{elia} M. Elia, M. Leone, {On the Inherent Space Complexity of
    Fast Parallel Multipliers for $GF(2^m)$}, {\em IEEE Trans. on
    Computer}, vol. 51, No. 3, March 2002, pp.346-351.
\bibitem{vetterli} J. Hong, M. Vetterli, Simple Algorithms for BCH
  Decoding, {\em IEEE Trans. on Communications}, Vol. 43, No. 8,
  August 1995, pp.2324-2333.
\bibitem{lidl} R. Lidl, H. Niederreiter, {\em Introduction to finite fields and their applications},
  Cambridge University Press, Cambridge, 1986.
\bibitem{knuth2}
     D.E. Knuth,
     {\em The Art of Computer Programming,}
     Seminumerical algorithms,
     vol.~II, Addison-Wesley, Reading Massachussetts, 1981.
\bibitem{sloane} F.J. MacWilliams, N.J.A. Sloane, {\it The Theory of
    Error-Correcting Codes}, North Holland, New York, 1977.
\bibitem{wicker2}
    S.B. Wicker, V.K. Bhargava, eds. {\em  Reed-Solomon codes and their applications}, IEEE Press, 
    Piscataway, N.J., 1994.
\end{thebibliography}
\end{document}